# DEVELOPMENT OF CLASSROOM MANAGEMENT BASED ON STUDENT LEARNING STYLE DATABASE

Raditya Bayu Rahadian [1(a)], C. Asri Budiningsih [2(b)]

[1] Department of Education Management, Yogyakarta State University, Indonesia
[2] Department of Instructional Technology, Yogyakarta State University, Indonesia

a) radityabayu.2020@student.uny.ac.id   b) asri_budi@uny.ac.id

| ARTICLE INFORMATIONS | ABSTRACT |
|---|---|
| **Keywords:**<br>classroom management, index of learning style, android app, instructional strategy, instructional media<br><br>**Corresponding Author:**<br>E-mail:<br>radityabayu.2020@student.uny.ac.id<br><br>DOI:<br>10.5281/zenodo.7618566<br><br>[CC BY-NC-ND] | *This study aims to produce a classroom management application based on a database of student learning styles to identify and analyze student learning styles and match them with appropriate learning methods and media. This application is expected to facilitate classroom management according to student preferences. This research was research and development. Respondents were 198 people consisting of 30 teachers and 168 students at a Junior High School in South Bangka, Indonesia. The data analysis used quantitative data analysis techniques using descriptive statistics. The results showed that the application developed was feasible to be used to identify and analyze student learning styles and provide suggestions for learning methods and media that teachers should use in classroom management. The level of application eligibility in terms of technical aspects reaches 4.75, and the content aspect reaches 4.80. This application proved to be useful for teachers in class management as evidenced by the achievement of the usability test score by the teacher reaching 4.40. The usefulness of the application from the student's perspective can be seen from the increase in the attractiveness of learning that occurs after the teacher uses this application compared to before.* |

## A. INTRODUCTION

Decisions in managing the class are often preoccupied with preparing so much material about what will be studied, then forgetting to consider how students can learn the material, the most effective approach to learning it, and the methods chosen to achieve learning objectives. Even students are often not made aware of their conditions, tendencies, and strengths in studying the material. Whereas the key to the efficiency and effectiveness of learning depends on concern for how students learn, which is closely related to the individual characteristics of student learning as the core of classroom management (Stobaugh, 2013: p.66).

Classroom management is a factor that can influence learning. This is due to the variety of components that teachers must pay attention to in the classroom, including factors related to students, methods, media, experiences, and conditions of learning facilities (Barringer, Pohlman, & Robinson, 2010; Pritchard, 2009; Reigeluth, Beatty, & Myers, 2017; Scheiter, Gerjets, Vollmann, & Catrambone, 2009; Willingham, Hughes, & Dobolyi, 2015).

The variety of learning components that the teacher must pay attention to (Stobaugh, 2013: p.59), the difficulty of adjusting the method and not choosing learning media according to student preferences (Brophy, Alleman, & Knighton, 2010; Duignan, 2006), inappropriate student grouping preferences (Samaras, Freese, Konsik, & Beck, 2008: p.158), the lack of technology that can help classroom management (Eady & Lockyer, 2013: p.91) causes the functions of classroom management unrun optimally.

Successful classroom management does not only make one instructional method the most effective method, but many methods will be the best methods for each material and student conditions and characteristics (Rosewell, 2005: p.1), due to basically someone has strength, character, and unique tendency to receive and process information as well as make it tend to prefer certain types of information in learning. This shows that everyone has a special way that involves a method or set of strategies in learning (Franzoni & Assar, 2009: p.18).

Student characteristics are believed to be a special ability that affects the degree of success in following a program (Barringer, Pohlman, & Robinson, 2010s: p.23). In another emphasis, Cronbach & Snow (1977) linked student characteristics as the key to interactions between students and learning that would affect the effectiveness of learning (Scheiter et al., 2009: p.388).

Learning styles can be explained as 1) a way a person learns; 2) the best or most preferred attitude (involving knowledge, and skills) in thinking, processing information, and learning performance; 3) habits, strategies, or mental behaviors that remain about learning, learning that is shown individually (Pritchard, 2009: p.41). Learning style is one of the characteristics of students which are included in the learning condition variables which are very important in their influence on learning (Felder & Brent, 2005: p.57; Mehlenbacher, 2010: p.221; Reigeluth, 1983: p.18; Scheiter, Gerjets, Vollmann, & Catrambone, 2009: p.388). In a somewhat different language, learning styles can be defined in two parts, namely as a different tendency to process certain types of information, or a tendency to process information in certain ways (Willingham, Hughes, & Dobolyi, 2015: p.266).

The learning style model used in this study is the Felder & Silverman learning style, with the following reasons:

1. One of the learning style models that has a strong influence on learning and learning design (Mehlenbacher, 2010: p.221).

2. It has been successfully implemented in previous research with regard to the adaptation of learning styles to the use of electronic learning materials and learning strategies (Dorça, Araújo, de Carvalho, Resende, & Cattelan, 2016; Franzoni & Assar, 2009; Sabine Graf, Lin, Kinshuk, & McGreal, 2012; Kowalski & Kowalski, 2013; Li, 2015; Popescu, Badica, & Moraret, 2010; Psycharis, Botsari, & Chatzarakis, 2014; Rajper, Shaikh, Shaikh, & Mallah, 2016).

3. The use of the model widely for educational and research purposes has been approved by its creators (Felder, 2002).

4. Easy to use even by students themselves and the results are easy to interpret for analysis (Franzoni & Assar, 2009; Ng, Pinto, & Williams, 2008).

5. Index of Learning Style has gone through long and in-depth validation so that it is appropriate to be used in defining student learning styles (Felder & Spurlin, 2005; S Graf, Viola S., Lea, & Kinshuk, 2007; Jingyun & Takahiko, 2015; Litzinger, Lee, Wise, & Felder, 2007; Ovariyanti & Santoso, 2017; Viola, Graf, Kinshuk, & Leo, 2006; Zywno, 2003).

The type of learning style influences the instructional method that should be chosen. The mismatch between the types of student learning styles and instructional methods will cause learning problems (Rahadian & Budiningsih, 2017: p.12). The various relationships between types of student learning styles and instructional methods are presented in Table 1.

**Table 1. Instructional Methods Fit the Type of Learning Style**

| Instructional Methods | Student Learning Styles | | | | | | | |
|---|---|---|---|---|---|---|---|---|
| | Active | Reflective | Sensitive | Intuitive | Visual | Verbal | Sequential | Global |
| Game & simulation | √ | | | √ | √ | | | |
| Problem solving based | √ | | √ | | | | | |
| Role playing | √ | | | √ | | | | √ |
| Student presentations | | √ | √ | | √ | | √ | |
| Panel discussions | √ | | | √ | | √ | | |
| Brainstorming | √ | | | | | √ | | |
| Case study | | √ | | √ | | | | √ |
| Question & answer | | √ | √ | | | √ | √ | |
| Project design | √ | | | √ | | | | √ |

(Source: Adapted from Franzoni & Assar, 2009; Rahadian & Budiningsih, 2017)

Table 1 shows that student learning style preferences affect the instructional methods to be used. For example, the game and simulation method is suitable for students with active, intuitive, and visual learning styles, but is less suitable for other learning styles. Learning styles not only affect the instructional method chosen, but also the instructional media that will be used. This is in accordance with the research that has been done which states that learning styles not only affect suitable instructional methods but also their suitability with the instructional media used (Rahadian & Budiningsih, 2017; Reigeluth et al., 2017). The instructional media used must be able to accommodate a learning style so that it can involve students directly in what is being learned. Students feel an attachment and an interest in what they are learning. The relationships between learning styles and instructional media are presented in Table 2.

**Table 2. Instructional Media Fit the Type of Learning Style**

| Instructional Media | Student Learning Styles | | | | | | | |
|---|---|---|---|---|---|---|---|---|
| | Active | Reflective | Sensitive | Intuitive | Visual | Verbal | Sequential | Global |
| Audio recording | | | | | | √ | √ | |
| Audio conference | | | | | | √ | √ | |
| Online community learning | | | | | √ | | | √ |
| Chatting/messenger | √ | | | | | | | √ |
| E-mail | √ | | | | | | | √ |
| Animation | | | | √ | √ | | | |
| Chart | | | | √ | √ | | | |
| Picture | | | | √ | √ | | | |
| Simulation | | | | | √ | | | |
| Magazine | | √ | | | | | √ | |
| Newspaper | | √ | | | | | | |
| Book/ebook | | √ | | | √ | | √ | |
| Hypertext/website | | √ | | | √ | | √ | |
| Slideshow | | √ | | √ | √ | | √ | |
| Internet research | √ | | | | √ | | | √ |
| Tutorial | | √ | | | | | | |
| Live video recording | | | | | √ | √ | | |
| Video conference | | | | | √ | √ | | |
| Video | | | | | √ | √ | | |

(Source: Adapted from Franzoni & Assar, 2009; Rahadian & Budiningsih, 2017)

Table 2 explains that students' learning styles also affect the tendency of instructional media they prefer and it means that they can learn the material well if it is presented with appropriate instructional media. For example, audio recording and audio conferencing are more suitable for intuitive and sequential learning styles, but should not be used for students with learning styles other than that.

Based on Table 1 and Table 2, it can be seen clearly that each type of learning style has a tendency for more appropriate instructional methods and media. This must be considered by teachers in classroom management. What will happen if the teacher manages the class not based on student learning styles? Students will have difficulty understanding the learning material, causing boredom, lack of student activity, low levels of understanding, and deteriorating quality and learning outcomes achieved (Budiningsih, 2011: pp.160-161).

Through the problems that have been raised, the author tries to answer them by developing a classroom management application, which is application that supports the spectrum of learning administration including classroom management (Churchill, Lu, Chiu, & Fox, 2016: p.13). The application uses a database of student learning styles as a starting point for classroom management which is realized by the android application as a user interface. This application can comprehensively help and support teachers in managing the classroom by making student learning styles the starting point in a series of responsible instructional development.

The user interface application was chosen based on Android mobile considering that the majority of teachers in South Bangka, Indonesia, use smartphones with the Android operating system. This learning style database-based classroom management application is expected to make it easier for teachers to manage classes based on appropriate student preferences and then adjust it to recommendations for appropriate learning methods and media. Teachers can identify groups of students based on their learning styles, so they can determine methods and provide appropriate learning media according to the group. This service will greatly save time, thought, cost, and teacher labor in managing the class well.

**B. MATERIAL AND METHOD**

The research method was research and development (R&D) which aims to produce a product in the form of a database application for learning styles for junior high school students based on Android. This application is an integrated Database Management System (DBMS) that consists of a database of student learning styles and an Android-based mobile device application to help teachers manage their classes. The product is developed to adopt the APPED development model which can be used as a reference in R&D research. The essence of this type of R&D is an element of research and development. The stages in the APPED model follow the logic of the type of R&D study, there are analysis and initial research, design, production, evaluation, and dissemination (Surjono, 2017: pp.66-74).

The research took place in South Bangka, Indonesia from February to June 2018. Development research respondents consisted of 9 (nine) subject teachers. The alpha test was carried out by 2 (two) media experts. The first beta test was conducted by 10 (ten) teachers. The second beta test was conducted by 21 (twenty-one) teachers. The assessment of the usefulness of the product involved 30 (thirty) teachers who were respondents. The assessment of learning style preferences and the attractiveness of learning involved 168 (one hundred and sixty-eight) students who were randomly selected as respondents who represented the class that was taught by the teacher who was the respondent in this development research.

The data in this research and development are qualitative and quantitative data. The data collection techniques used in this study were questionnaires, interviews, and observations. The instruments used in this research and development consist of a needs questionnaire, a product feasibility questionnaire, an Index of Learning Styles (ILS) questionnaire, a product usefulness questionnaire, a learning attractiveness questionnaire, an interview guide, an observation sheet, and documentation.

Qualitative data were obtained from the results of the needs analysis through observations and interviews conducted at the needs analysis stage and initial research. Quantitative data were obtained from needs analysis, media expert assessment, and teacher and student questionnaires which were analyzed descriptively using a Likert scale. These data are described as material for consideration for revising the development of the classroom management application based on the student learning style database. In addition, quantitative data is also used to measure the quality of the usefulness of the products developed so that they can be used in classroom management.

The data analysis technique used in this research is quantitative using descriptive statistics. Descriptive statistical analysis is the statistic used to analyze data by describing the collected data as it is. However, descriptive statistical analysis can still present data and explain it as a justifiable conclusion, even though it is not intended to generalize (Cohen, Manion, & Morrison, 2007:pp.503-504). The steps in quantitative data analysis are scaling obtained from a questionnaire using a Likert scale. For the product feasibility questionnaire using a score: very feasible = 5, feasible = 4, quite feasible = 3, less feasible = 2, not feasible = 1. For the product usefulness questionnaire using a score: strongly agree = 5, agree = 4, doubtful = 3, disagree = 2, strongly disagree = 1 (Best & Kahn, 2006:p.331).

## C. RESULTS

Initial product development involving prospective users is intended to produce database-based classroom management applications for student learning styles that suit the needs of prospective users. In addition, literature studies are also used to translate the interest of potential users into a product that is not only in accordance with needs but in accordance with the underlying theory.

Based on the design results, this product aims to provide services to identify student learning styles and provide adaptive advice regarding the implementation of instructional strategies and media that can be used as material for consideration in classroom management. The content of this product includes a database of student learning styles in terms of the Felder-Silverman learning style model equipped with an Index of Learning Styles (ILS) questionnaire to identify types of student learning styles, management of students in study groups, as well as adaptive suggestions related to instructional strategies and media. which corresponds to the type of learning style detected.

Based on the design of product use procedures, there are several menus and services that must be in the application, namely the login menu, student management, group management, ILS questionnaire filling services, as well as information on strategic and media suggestions that are in accordance with the type of student learning style. This initial menu is then used to develop an application scheme that accommodates the services required by potential users. The relationship scheme between menus and sub-menus are shown in Figure 1.

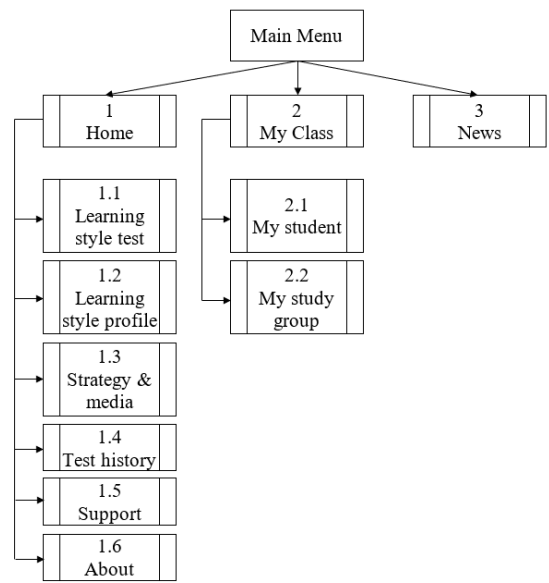

**Figure 1. Main Menu and Sub Menu Relationship Scheme**

Figure 1 shows the relationship between the menus in the application. Menu grouping is done based on general and specific activities. The sub-menu on the 'Home' menu is more general in nature and requires a few user settings. In the 'Home' menu, there will be several sub-menus, such as learning style tests, learning style profiles, strategies and media, test history, support, and about, as shown in Figure 2. The learning style test menu contains the ILS questionnaire to determine the type of student learning style. The learning style profile menu contains complete information on all types of learning styles in the Felder-Silverman model. The strategy and media menu contains information about strategies and media that can be used in learning. The test history menu contains all the learning style test data that students have done. The support menu contains general information on the Felder-Silverman model learning style. The "about" menu contains information about applications and developers.

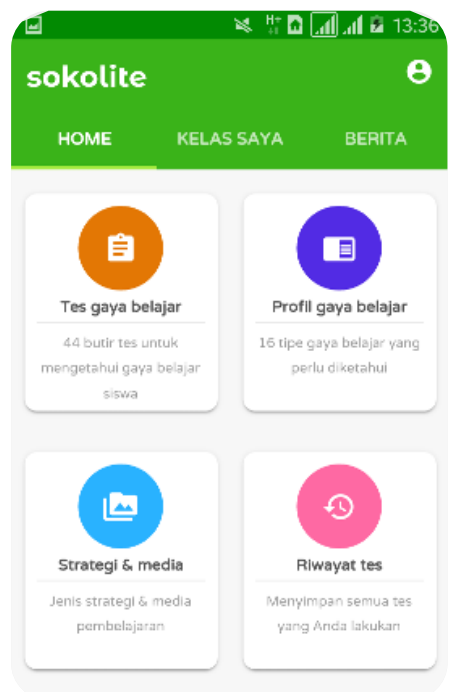

**Figure 2. Display of the "Home" Menu**

The sub-menu on the 'My Class' menu is more specific, where each user can make flexible arrangements regarding the student and study group that the user is assisting, as shown in Figure 3. The 'News' menu is a menu that is planned for further development which is not included in this research.

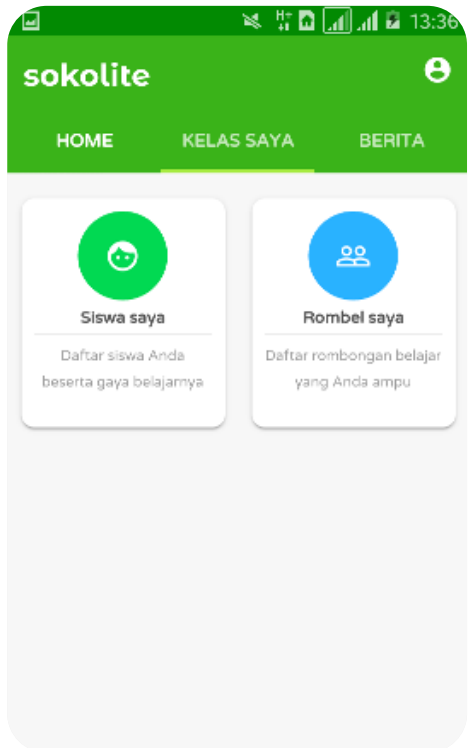

**Figure 3. Display of the "My Class" Menu**

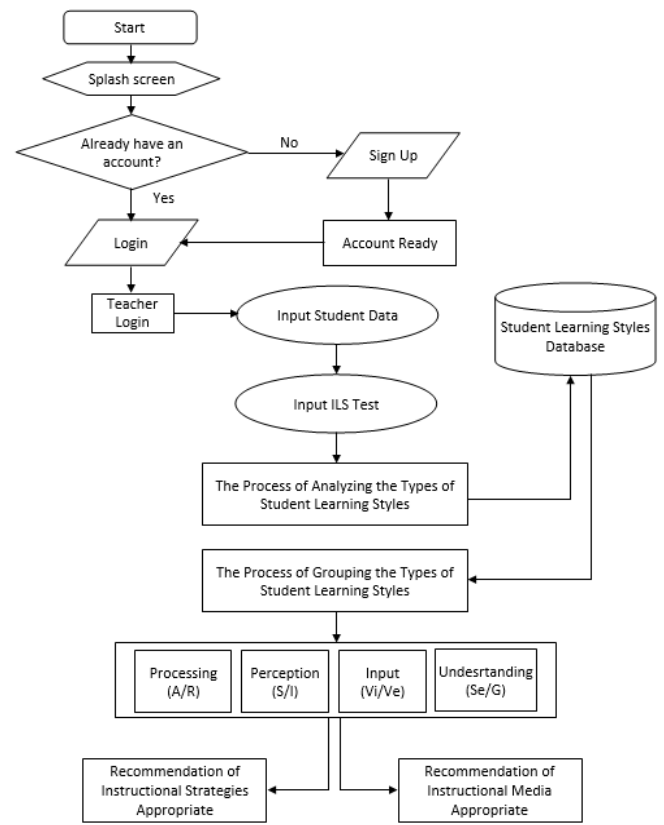

**Figure 4. Flow of Learning Style Identification**

This product was developed to overcome problems that occur in classroom management where teachers have difficulty determining strategies and learning media that suit the needs of students, especially their learning styles so classroom management is not effective. Therefore the product must be able to assist teachers in identifying and analyzing student learning styles before determining appropriate learning strategies and media. The flow of the process of identifying and analyzing student learning styles in the Android application can be described in Figure 4.

Based on Figure 4, it can be seen that the user simply enters the results of the ILS questionnaire that has been filled in by the student into the application, then the system will process it automatically, save the input data and identification data into the database. In the following process, the system will provide the results of the type of learning style detected and provide recommendations regarding the selection of strategies and instructional media that are in accordance with the kind of learning style detected. This will greatly facilitate the teacher so that there is no need to bother doing manual calculations which are tiring and have the potential to cause miscalculations so that the recommendations given will be inappropriate.

The student learning style database that has been collected can be seen by students and by class. The students' ILS test results will be stored and can be viewed individually or in groups, as shown in Figure 5.

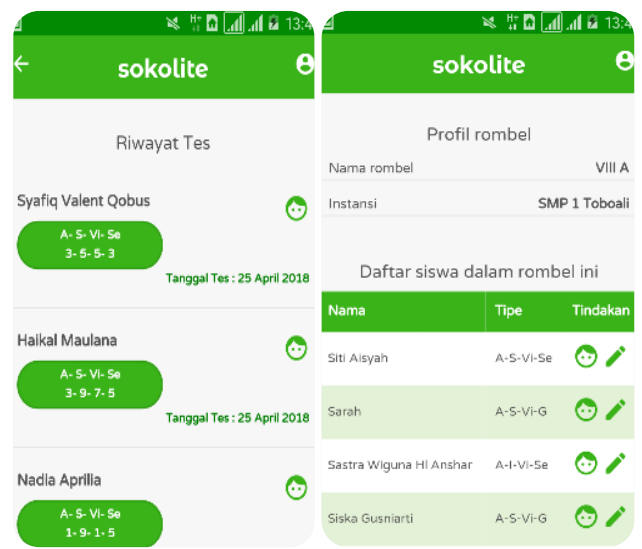

**Figure 5. Student ILS Test Results**

The product developed refers to the eligibility criteria for the technical aspects of an Android application and the content aspects as a guide in identifying student learning styles accurately. Based on the second beta test, it can be concluded that the product is very feasible in terms of technical and content aspects. The feasibility level of the product in terms of technical aspects reaches 4.75, and the aspect of content reaches 4.80. The product proved to be useful for the teacher in planning the lesson as evidenced by the achievement of the product usefulness test score by the teacher reaching 4.40. The usefulness of the product from the student's point of view can be seen from the level of attractiveness of learning that takes place after the teacher uses the product, which is much higher than the level of attractiveness of learning before the teacher uses the product.

The product developed must meet a good level of usefulness in learning. The level of usefulness of the product can be seen from the point of view of teachers and students. The usefulness of the product from the teacher's point of view can be seen from the achievement of high scores on the product usefulness questionnaire by the teacher. The usefulness of the product from the student's point of view can be seen from the level of attractiveness of learning that takes place after the teacher uses the product, which is much higher than the level of attractiveness of learning before the teacher uses the product.

## D. Discussion

The product developed is a classroom management application based on a database of student learning styles, which is a single DBMS consisting of a student learning style database and an Android-based mobile device application. A real-time database that contains data on student learning styles that can be accessed online. This product can be accessed via a mobile device with a minimum specification Android operating system version 4.0 (Ice Cream Sandwich), 1GB RAM, a touch screen with a resolution of 320x240 pixels, and an active internet connection.

The product developed refers to the achievement of feasibility standards in the technical aspects of Android application development, as well as the content aspects based on the Felder-Silverman learning style model. The trial in stages starts from expert validation, beta I test, and beta II tests based on predetermined standard eligibility criteria guaranteeing that the resulting product can meet an adequate level of feasibility for use in assisting learning planning. Judging by the achievement of the score on the technical aspect which reaches 4.75, and the content aspect is 4.80, it can be concluded that the product developed is very suitable to be used to assist learning planning.

This application supports the success of classroom management. Successful classroom management does not only make one instructional method the most effective method, but many methods will be the best methods for each material and student conditions and characteristics (Rosewell, 2005:p.1), due to basically someone has strength, character, and unique tendency to receive and process information as well as make it tend to prefer certain types of information in learning. This shows that everyone has a special way that involves a method or set of strategies in learning (Franzoni & Assar, 2009:p.18).

Classroom management is a factor that can influence learning. This is due to the variety of components that teachers must pay attention to in the classroom, including factors related to students, methods, media, experiences, and conditions of learning facilities (Barringer et al., 2010; Pritchard, 2009; Reigeluth et al., 2017; Scheiter et al., 2009; Willingham et al., 2015). This product makes it easy for teachers to get information on student learning styles quickly and easily. The teacher only needs to enter the results of the student's questionnaire and then an explanation will automatically appear regarding the types of student learning styles as well as media recommendations and learning strategies that are in accordance with the type of student's learning style (Felder & Brent, 2005; Rahadian & Budiningsih, 2017).

Recommendations are also given at the learning group level, where the total types of student learning styles in one group are analyzed collectively to determine the appropriate media recommendations and learning strategies for students in a particular group (Churchill et al., 2016:p.13). The application uses a database of student learning styles as a starting point for classroom management which is realized by the android application as a user interface. This application can comprehensively help and support teachers in managing the classroom by making student learning styles the starting point in a series of responsible instructional development. Student data along with information on the type of learning style are stored in a database that can be accessed by all users in real time. The teacher can search the name of the student so that if the student is already in the database, the teacher can immediately add them as a student without having to re-enter the student's data and their learning styles. This service will save teachers more time in identifying types of student learning styles.

## E. Conclussion

The application developed provides a complete service that makes it easier for teachers to manage their classes. The learning style model used in this application has comprehensively described the characteristics of student learning styles from various dimensions (processing, perception, input, and understanding) related to the selection of strategies and learning media in classroom management. Teachers can use the product to identify and analyze student learning styles individually and collectively (in groups) to obtain appropriate suggestions regarding the strategies and learning media to be used. In addition, information on the types of student learning styles can be conveyed to students and parents so that they know the tendency of student learning styles and can provide appropriate assistance/facilities to support student learning more effectively. This study only involved teachers and junior high school students as respondents. Further research involving respondents at various levels should be carried out to ensure the effectiveness of using this application at various levels.

## F. Aknlowledgment

The authors would like to thank Prof. Herman Dwi Surjono, M.T., M.Sc., Ph.D., and Prof. Dr. Herminarto Sofyan, M.Pd. as expert validators. In addition, the author is also grateful to Prof. K.H. Sugijarto, M.Sc., Ph.D., for the review and analysis of this article.